\documentclass[reprint,aip,jcp]{revtex4-1}

\usepackage{graphicx}
\usepackage{dcolumn}
\usepackage{bm}
\usepackage{amsmath}
\usepackage{lipsum}
\usepackage{color}

\newcommand{\eg}{\mbox{e.\,g.\,}\ }
\newcommand{\ie}{\mbox{i.\,e.\,}\ }

\begin{document}

\title{Chain End Mobilities in Polymer Melts -- A Computational Study}

\author{Diddo Diddens}
\affiliation{University of M\"unster, Institute of Physical Chemistry, Corrensstrasse 28/30, 48149 M\"unster, Germany}
\email[Current address: Institut Charles Sadron, Universit\'e de Strasbourg, 23~Rue du Loess, BP~84047, 67034~Strasbourg, France, \\ Electronic mail: ]{diddo.diddens@ics-cnrs.unistra.fr}
\author{Andreas Heuer}
\affiliation{University of M\"unster, Institute of Physical Chemistry, Corrensstrasse 28/30, 48149 M\"unster, Germany}

\date{\today}

\begin{abstract}
The Rouse model can be regarded as the standard model to describe the dynamics of a short polymer chain under melt conditions. 
In this contribution, we explicitly check one of the fundamental assumptions of this model, namely that of a uniform friction coefficient for all monomers, on the basis of MD simulation data of a poly(ethylene oxide) (PEO) melt. 
This question immediately arises from the fact that in a real polymer melt the terminal monomers have on average more intermolecular neighbors than the central monomers, and one would expect that exactly 
these details affect the precise value of the friction coefficient. 
The mobilities are determined by our recently developed statistical method, which provides detailed insights about the local polymer dynamics. 
Moreover, it yields complementary information to that obtained from the mean square displacement (MSD) or the Rouse mode analysis. 
It turns out that the Rouse assumption of a uniform mobility is fulfilled to a good approximation for the PEO melt. 
However, a more detailed analysis reveals that the underlying microscopic dynamics is highly affected by different contributions from intra- and 
intermolecular excluded volume interactions, which cannot be taken into account by a modified friction coefficient. 
Minor deviations occur only for the terminal monomers on larger time scales, which can be attributed to the presence of two different escape mechanisms from their first coordination sphere. 
These effects remain elusive when studying the dynamics with the MSD only. 
\end{abstract}


\maketitle

\section{Introduction}

The Rouse model~\cite{Rouse,DoiEdwards} is one of the standard models to describe the dynamics in a polymer melt of non-entangled 
chains. Here, the polymer chain is modeled as a sequence of $N$ harmonically linked beads. 
All intermolecular interactions of the chain are reduced to a frictional and a stochastic force, both characterized by the friction 
coefficient $\zeta$ via the Fluctuation-Dissipation Theorem~\cite{DoiEdwards}. 
Within this model, $\zeta$ takes the same value for all beads irrespective of the monomer position~$n$. 
However, when going to real polymer melts, it is obvious that the individual polymer segments along the chain cannot have exactly 
the same intermolecular environment due to chain connectivity.  
A terminal segment has on average more intermolecular neighbors in its first coordination sphere than a segment located in 
the center of the chain. 
Therefore, it is questionable if the assumption of a uniform friction constant along the entire chain as assumed in the Rouse model 
still holds for a realistic chain in the melt. 

The dynamics of the chain ends also plays a special role in the limit of long chains, which is usually described by the reptation 
model~\cite{deGennes}. 
Within this picture, the topological constraints imposed by the other chains are modeled as a tube~\cite{DoiEdwards}, and the tagged 
chain performs Rouse-like motion within this effective tube. 
Here, the local dynamics of the chain ends play an important role in a twofold manner: 
First, since the chain can only escape the tube at its ends, one would expect that the dynamics of the chain ends significantly 
influences the overall relaxation mechanism. 
For example, it has been observed in simulations~\cite{KremerJCP1990,BaschnagelBinderMamol2001,PaulChemPhys2002} that the tube 
constraints are less pronounced for the terminal monomers. 
More recently, it has been shown~\cite{GrestEPL2011} that exactly this effect also plays an important role in welding processes of 
two entangled polymer films brought into contact with each other, as the initial interdiffusion across the film-film interface is 
highly governed by the faster chain ends. 
Improved models for entangled polymer melts suggest that the tube relaxation is enhanced by so-called contour-length fluctuations~\cite{DoiEdwards}, 
which have also been observed experimentally~\cite{RichterEPL2005}. 
Here, the motion of the chain ends leads to a loss of memory of the initial tube. 
Second, switching to the surrounding chains imposing the tube constraint, it was argued that the chain ends do not contribute to the 
formation of entanglements, and a revised tube model has been proposed~\cite{KavalassisPRL1987,KavalassisMamol1988}. 
Experimental data of a bimodal melt was successfully interpreted within this concept~\cite{RichterJCP1999}. 

A different but related interplay of topological constraints and chain end dynamics becomes important if a polymer melt approaches 
the glass transition temperature $T_\mathrm{g}$. 
The dependence of $T_\mathrm{g}$ on the molecular weight is usually described by the empirical Flory-Fox equation~\cite{FoxFloryJApplPhys1950}. 
Within this picture, the chain ends experience a larger amount of free volume than the central monomers, leading to a lower $T_\mathrm{g}$ for 
short chains, which has also been observed in MD simulations~\cite{BaschnagelEPJE2011}. 

For all these reasons, the specific role of the chain-end dynamics as compared to the motion of the central parts of the chain is a long-standing 
issue in polymer science, which has naturally already been investigated by several experimental~\cite{KitaharaPolymer1980,WelpMamol1999,MiwaMamol2005,MiwaMamol2010} 
and numerical~\cite{KremerJCP1990,BaschnagelBinderMamol2001,PaulChemPhys2002,PaulSmithRev2004,VogelMamol2010} studies. 
However, when discussing \lq segmental mobilities\rq, one should keep in mind that this expression is used ambiguously in the literature: 
In its most general meaning, the term refers to the overall motion of a given segment, which for longer time scales is not only determined by the \emph{bare} 
Rouse mobility $\zeta^{-1}$, but also by the chain connectivity, which is naturally less pronounced for terminal monomers. 
Within this context, it is important to mention that most experiments (such as spin-labeling techniques~\cite{KitaharaPolymer1980,MiwaMamol2005,MiwaMamol2010} 
or neutron reflectivity~\cite{WelpMamol1999}/neutron spin echo~\cite{RichterEPL2005,RichterJCP1999} measurements) as well as standard observables calculated 
from simulation data (such as the mean square displacement~\cite{KremerJCP1990,BaschnagelBinderMamol2001,PaulChemPhys2002,PaulSmithRev2004}) quantify the net 
movement of a polymer segment, and thus rather measure an interplay between bare mobility and chain connectivity. 

In contrast to this, we focus on the microscopic Rouse mobility $\zeta^{-1}$ \emph{free} from connectivity effects. 
In particular, we check if the classical Rouse assumption of a uniform $\zeta$-value for all polymer segments, independent of their position 
within the chain (\ie at the end or in the center), is fulfilled for a chemically realistic chain under melt conditions. 
This is a highly pertinent question, since it is obvious that the intermolecular environment, which determines the precise value of 
$\zeta$ within the picture of Brownian motion~\cite{DoiEdwards}, is significantly different for a terminal and a central monomer. 
In order to address this issue, we apply our recently developed method to determine segmental mobilities (pq-method)~\cite{DiddensEPL2010} to 
atomistic MD simulation data of a poly(ethylene oxide) (PEO) melt. 
This approach has been devised to extract specific information about the {\it local} friction. 
The results from the MD simulations are interpreted within the simpler semiflexible chain model (SFCM)~\cite{WinklerJCP1994}, in which 
a Rouse chain is augmented by an additional angle potential, thus incorporating chain stiffness. 

From a conceptual point of view, we put additional emphasis on two points: 
First, to highlight the range of capabilities of our method, we also apply it to Brownian dynamics simulations for a simple polymer model 
and second, we discuss its information content in comparison to the standard analysis of the mean square displacement (MSD).

\section{Simulation Details}
\label{sec:sim_details}

For our analysis, we used MD simulation data of a PEO melt from a previous study~\cite{MaitraHeuerMCP2007}. 
Here, the simulation cell consisted of $16$ PEO chains with $N=48$ monomers each. 
The simulations were performed in the $NVT$ ensemble with the GROMACS simulation package~\cite{GROMACS} using an effective two-body 
polarizable force field~\cite{BorodinSmithJPCB2003}. 
The temperature had been maintained at $T=450\text{~K}$ by a Nos\'{e}-Hoover thermostat. 
Further technical details can be found in the original study~\cite{MaitraHeuerMCP2007}. 

Additionally, we simulated our reference, \ie the SFCM~\cite{WinklerJCP1994}, via Brownian Dynamics (BD) simulations. 
In this model, a Rouse chain is augmented by a bending potential of the form 
\begin{equation}
 \label{eq:angle_pot}
 \begin{aligned}
  U_\theta(\lbrace{\bf R}_n\rbrace) &= -\lambda\sum_{n=2}^{N-1}\frac{({\bf R}_{n+1}-{\bf R}_{n})\cdot({\bf R}_{n}-{\bf R}_{n-1})}{|{\bf R}_{n+1}-{\bf R}_{n}||{\bf R}_{n}-{\bf R}_{n-1}|} \\
                                    &= -\lambda\,\sum_{n=2}^{N-1}\cos{\theta_n}\,\text{,}
 \end{aligned}
\end{equation}
where the ${\bf R}_{n}$ correspond to the position vectors of the beads ($n=1,2, \, \ldots, \, N$), and $\theta_n$ is the bonding angle 
defined by beads $n-1$, $n$ and $n+1$. 
In our present work, we chose a value of $\beta\lambda=2.1$ (with $\beta^{-1}=k_\mathrm{B}T$ 
being the thermal energy), for which we found that the characteristic ratio $C_\infty$ 
of the SFCM matches that of the PEO chains in the MD simulations ($C_\infty\approx 3.2$). 
The semiflexible model chains contained the same number of monomers as the PEO chains ($N=48$). 
For the friction coefficient $\zeta$, the temperature $k_\mathrm{B}T$ and the mean squared bond 
length $b_0^2$, unit values have been used. 
The model polymers were propagated by a simple Euler integrator using an elementary time step 
of $\Delta t=0.0001$.

\section{pq-Method}

Although the pq-method has already been described previously~\cite{DiddensEPL2010}, 
we nevertheless begin with a systematic review. 
This is mainly due to the fact that we complement our numerical scheme by additional analytical 
calculations in the present article. 

\begin{figure}
 \centering
 \includegraphics[scale=0.175]{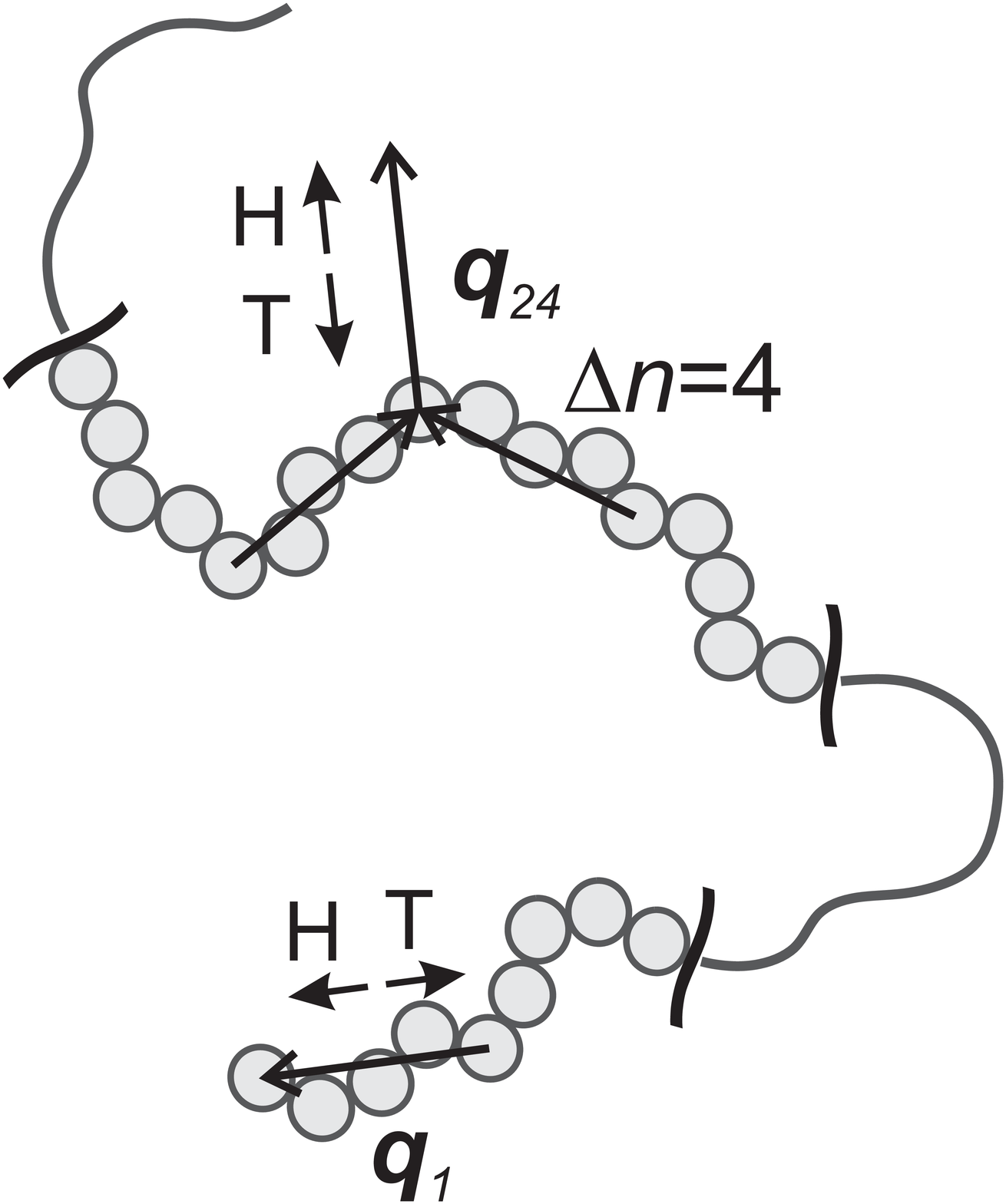}
 \caption{Sketch depicting the definitions used in the present analysis for a simplified excluded volume chain. 
          Depending on the direction in which the center of mass of the subchain 
          consisting of all monomers within ${\bf q}_n$ moves during time $\Delta t$, 
          one can distinguish between head (H) and tail (T) monomers (see text for further explanation). }
 \label{fig:sketch}
\end{figure}

The basic idea of our approach is a stroboscopic view on the local Langevin dynamics of 
monomer $n$ of a Rouse-like polymer chain, characterized by the equation 
\begin{equation}
 \label{eq:langevin}
 {\bf p}_{n}(t,\Delta t)=-A_n(\Delta t)\,{\bf q}_{n}(t)+\mathrm{noise}\,\text{,}
\end{equation} 
where the definitions 
\begin{equation}
 {\bf p}_n(t,\Delta t)=(1/\Delta t)\left[{\bf R}_n(t+\Delta t)-{\bf R}_n(t)\right]
\end{equation}
and 
\begin{equation}
 \label{eq:q_def}
 {\bf q}_n(t)=2\,{\bf R}_n(t)-{\bf R}_{n+\Delta n}(t)-{\bf R}_{n-\Delta n}(t)
\end{equation}
have been used. 
In case of the terminal monomers, Eq.~\ref{eq:q_def} can easily be modified according to 
${\bf q}_1={\bf R}_1-{\bf R}_{1+\Delta n}$ and ${\bf q}_N={\bf R}_N-{\bf R}_{N-\Delta n}$ 
(see sketch in Fig.~\ref{fig:sketch}). 
Here, the parameter $\Delta n$ is in practice determined by the local chain stiffness, 
which will be discussed subsequently. 

For a Rouse chain ($\Delta n=1$), the quantity $A_n(\Delta t)$ in Eq.~\ref{eq:langevin} 
takes the value $k/\zeta$ in the limit $\Delta t\rightarrow 0$, where $k=3\,(\beta\,b^2)^{-1}$ 
is the entropic force constant, characterized by the average squared size $b^2$ of the Rouse segment. 
This is due to the fact that for this particular case, it is possible to interpret Eq.~\ref{eq:langevin} 
as the discretized Langevin equation of the Rouse model. 
For larger $\Delta t$, $A_n(\Delta t)$ naturally decreases, as the dynamics of bead $n$ 
also becomes affected by the motion of more remote beads, which cause additional backdragging forces. 
Note that for finite $\Delta t$, also the second term in Eq.~\ref{eq:langevin} can be still interpreted as a random term, 
at least if averaged over all possible chain configurations, compatible with a given ${\bf q}_n(t)$. 

For polymer chains with a certain stiffness, the parameter $\Delta n$ in Eq.~\ref{eq:langevin} 
has to be determined to facilitate comparison with the Rouse picture. 
In particular, it has to be assured that the adjacent bond vectors ${\bf R}_{n}-{\bf R}_{n-\Delta n}$ 
and ${\bf R}_{n+\Delta n}-{\bf R}_{n}$ are roughly independent from each other in their orientations. 
Thus, $\Delta n$ corresponds to the number of chemical monomers within one Kuhn segment. 
We chose $\Delta n$ such that 
$C_\infty \approx [\langle({\bf R}_{n}-{\bf R}_{n-\Delta n})^2\rangle/\langle({\bf R}_{n}-{\bf R}_{n-1})^2\rangle]^{1/2}$ 
is approximately fulfilled. 
The characteristic ratio in turn was determined from the mean squared end-to-end vector 
$\langle{\bf R}_\mathrm{e}^2\rangle$ using the identity $C_\infty=\langle{\bf R}_\mathrm{e}^2\rangle/[(N-1)b_0^2]$. 
Due to the choice of $\lambda$ in Eq.~\ref{eq:angle_pot}, the same $C_\infty$-value as for the PEO 
($C_\infty\approx 3.2$) chains was obtained for the SFCM, yielding $\Delta n=4$ for both systems.

\subsection{Analytical Calculation}
\label{ssec:analytical}

In our previous work~\cite{DiddensEPL2010,DiddensEPL2011}, we treated Eq.~\ref{eq:langevin} as a 
linear regression problem. 
According to this interpretation, $A_n(\Delta t)$ is simply given by 
$A_n(\Delta t)=\langle{\bf p}_n(\Delta t)\cdot{\bf q}_n\rangle/\langle{\bf q}_n^2\rangle$. 
Whereas $\langle{\bf q}_n^2\rangle$ is a trivial normalization factor, all dynamical information is contained in the correlator 
$\langle{\bf p}_n\cdot{\bf q}_n\rangle = \langle{p}_{n,x}{q}_{n,x}\rangle + \langle{p}_{n,y}{q}_{n,y}\rangle + \langle{p}_{n,z}{q}_{n,z}\rangle$. 
We start by calculating $\langle{\bf p}_n\cdot{\bf q}_n\rangle$ for bead $n$ of a semiflexible Rouse chain with an \emph{arbitrary} angle potential 
of the form $U_{\theta_n} = -\,\lambda\,F(\cos\theta_n)$ in the limit $\Delta t\rightarrow 0$. 
Within our subsequent calculus, we will first restrict ourselves to the case $\Delta n=1$. 
However, as we will argue below, the same value for $\langle{\bf p}_n\cdot{\bf q}_n\rangle$ 
is obtained for larger $\Delta n$. 
Furthermore, we assume that $k$ has a uniform value, while the friction coefficient may 
depend on $n$, \ie $\zeta=\zeta_n$ 
(note that the choice of a uniform $k$ for PEO is supported by the MD data, since we observe 
an identical distribution function for all Kuhn bonds ${\bf R}_{n}-{\bf R}_{n-\Delta n}$). 

As a starting point of our calculus, we separate ${\bf p}_n\cdot{\bf q}_n$ into its individual 
contributions, \ie 
\begin{equation}
 {\bf p}_n\cdot{\bf q}_n={\bf p}_{n,\mathrm{bond}}\cdot{\bf q}_n+{\bf p}_{n,\mathrm{bend}}\cdot{\bf q}_n+{\bf p}_{n,\mathrm{noise}}\cdot{\bf q}_n\text{,}
\end{equation}
where ${\bf p}_{n,\mathrm{bond}}$, ${\bf p}_{n,\mathrm{bend}}$ and ${\bf p}_{n,\mathrm{noise}}$ 
arise form the harmonic bonds, the stiffness potential and the fluctuating force, respectively. 
For an elementary time step in the BD simulations, we have according to the equations of motion 
\begin{equation}
 \label{eq:pbondq}
 \begin{aligned}
  {\bf p}_{n,\mathrm{bond}}\cdot{\bf q}_n &= \frac{k}{\zeta_n}({\bf r}_{n}-{\bf r}_{n-1})\cdot({\bf r}_{n-1}-{\bf r}_{n}) \\
                                          &= \frac{k}{\zeta_n}\left[-(r_{n-1}^2+r_{n}^2)+2\,r_{n-1}\,r_{n}\,(\hat{\bf r}_{n-1}\cdot\hat{\bf r}_{n})\right]
 \end{aligned}
\end{equation}
and 
\begin{equation}
 \label{eq:pbendq}
 \begin{aligned}
  {\bf p}_{n,\mathrm{bend}}\cdot{\bf q}_n &= &\frac{\lambda}{\zeta_n} \left[-\nabla_{n}F(\cos\theta_n)\right]\cdot({\bf r}_{n-1}-{\bf r}_{n}) \\
                                          &= &\frac{\lambda}{\zeta_n} \frac{\partial F(\cos\theta_n)}{\partial (\cos\theta_n)}\left[\frac{r_{n}}{r_{n-1}}+\frac{r_{n-1}}{r_n}\right] \\ 
                                          &  &\times \left[(\hat{\bf r}_{n-1}\cdot\hat{\bf r}_n)^2-1\right]\text{~,}
 \end{aligned}
\end{equation}
where we have used the notation ${\bf r}_{n-1}={\bf R}_{n}-{\bf R}_{n-1}$ and ${\bf r}_{n}={\bf R}_{n+1}-{\bf R}_{n}$ 
for the two bond vectors entering ${\bf q}_n$, $r_{n-1}$ and $r_{n}$ denote their respective lengths, and the 
hats indicate unit vectors. 
The Hamiltonian of the entire chain is given by 
\begin{equation}
 \label{eq:hamiltonian}
 H = \frac{k}{2}\sum_{n=1}^{N-1}\,{\bf r}_{n}^2 - \lambda\sum_{n=2}^{N-1}\,F(\hat{\bf r}_{n-1}\cdot\hat{\bf r}_{n})\text{~.}
\end{equation}
Thus, the expectation value $\langle {\bf p}_n\cdot{\bf q}_n\rangle$ is given by 
\begin{equation}
 \label{eq:pq_partition}
 \begin{aligned}
    &\langle {\bf p}_n\cdot{\bf q}_n\rangle = \langle {\bf p}_{n,\mathrm{bond}}\cdot{\bf q}_n\rangle + \langle {\bf p}_{n,\mathrm{bend}}\cdot{\bf q}_n\rangle \\
    &= \frac{k}{\zeta_n}\,\frac{\Pi_{m=1}^{N-1}\int_{-\infty}^{\infty}d\,{\bf r}_{m}\,{\bf p}_{n,\mathrm{bond}}\cdot{\bf q}_n\,\exp{(-\beta\,H)}}{\Pi_{m=1}^{N-1}\int_{-\infty}^{\infty}d\,{\bf r}_{m}\,\exp{(-\beta\,H)}} \\
    &+ \frac{\lambda}{\zeta_n}\,\frac{\Pi_{m=1}^{N-1}\int_{-\infty}^{\infty}d\,{\bf r}_{m}\,{\bf p}_{n,\mathrm{bend}}\cdot{\bf q}_n\,\exp{(-\beta\,H)}}{\Pi_{m=1}^{N-1}\int_{-\infty}^{\infty}d\,{\bf r}_{m}\,\exp{(-\beta\,H)}}
 \end{aligned}
\end{equation}
since $\langle{\bf p}_{n,\mathrm{noise}}\cdot{\bf q}_n\rangle = 0$. 
Interestingly, Eqs.~\ref{eq:pbondq}, \ref{eq:pbendq} and \ref{eq:hamiltonian} only depend 
on the relative orientation of ${\bf r}_{n-1}$ and ${\bf r}_{n}$ as well as their respective lengths. 
Thus, when calculating the thermodynamic average $\langle{\bf p}_n\cdot{\bf q}_n\rangle$, 
one can average over all possible orientations of a given bond vector ${\bf r}_{m}$ relative 
to the adjacent bond vector ${\bf r}_{m-1}$ in chain-internal spherical coordinates, 
\ie $\int_{-\infty}^{\infty}d\,{\bf r}_{m}\rightarrow\int_{0}^{\infty}d\,r_{m}\,r_{m}^2\int_{0}^{\pi}d\,\theta_m\,\sin\theta_m\int_{0}^{2\pi}d\,\phi_m$. 
Finally, one can average over all possible orientations of the last bond vector ${\bf r}_{1}$ 
in absolute space. 
The Gaussian integrals for the squared bond lengths $r_{n}^2$ can now easily be solved. 
For the integrals over the bonding and the bending forces (Eq.~\ref{eq:pq_partition}), 
denoted as $I_\mathrm{bond}$ and $I_\mathrm{bend}$ in the following, as well as for the 
partition function $Z$ in the denominator of Eq.~\ref{eq:pq_partition}, this leads to 
\begin{widetext}
  \begin{equation}
    \label{eq:bondint}
    \begin{aligned}
      I_\mathrm{bond} &= \Pi_{m=1}^{N-1}\int_{0}^{\infty}d\,r_{m}\,r_{m}^2\int_{0}^{\pi}d\,\theta_m\,\sin\theta_m\int_{0}^{2\pi}d\,\phi_m\,\left[-(r_{n-1}^2+r_{n}^2)+2\,r_{n-1}\,r_{n}\,(\hat{\bf r}_{n-1}\cdot\hat{\bf r}_{n})\right]\,\exp{(-\beta\,H)} \\
                      &= 2\,(2\pi)^2\,(\beta k)^{-4}\,\left[(2\pi^3(\beta k)^{-3})^{1/2}\,I_0\right]^{N-3}\,\int_{-1}^{1}d\,u_n\,\left[-3\pi+8u_n\right]\,\exp{\left(\beta\lambda\,F(u_n)\right)}\text{~,}
    \end{aligned}
  \end{equation}
  \begin{equation}
    \label{eq:bendint}
    \begin{aligned}
      I_\mathrm{bend} &= \Pi_{m=1}^{N-1}\int_{0}^{\infty}d\,r_{m}\,r_{m}^2\int_{0}^{\pi}d\,\theta_m\,\sin\theta_m\int_{0}^{2\pi}d\,\phi_m\,\frac{\partial F(\cos\theta_n)}{\partial (\cos\theta_n)}\left[\frac{r_{n}}{r_{n-1}}+\frac{r_{n-1}}{r_n}\right]\left[\cos^2\theta_n-1\right]\,\exp{(-\beta\,H)} \\
                      &= 8\,(2\pi)^2\,(\beta k)^{-3}\,\left[(2\pi^3(\beta k)^{-3})^{1/2}\,I_0\right]^{N-3}\,\int_{-1}^{1}d\,u_n\,\left[u_n^2-1\right]\,\frac{\partial F(u_n)}{\partial u_n}\,\exp{\left(\beta\lambda\,F(u_n)\right)}\text{~,}
    \end{aligned}
  \end{equation}
  and 
  \begin{equation}
    \label{eq:partfunc}
    \begin{aligned}
      Z &= \Pi_{m=1}^{N-1}\int_{0}^{\infty}d\,r_{m}\,r_{m}^2\int_{0}^{\pi}d\,\theta_m\,\sin\theta_m\int_{0}^{2\pi}d\,\phi_m\,\exp{(-\beta\,H)} \\
        &= 2\,\left[(2\pi^3(\beta k)^{-3})^{1/2}\right]^{N-1}\,I_0^{N-2}\text{~,}
    \end{aligned}
  \end{equation}
\end{widetext}
where $I_0=\int_{-1}^{1}d\,u_m\,\exp{\left(\beta\lambda\,F(u_m)\right)}$ defines the integral 
over any bonding angle with $m\neq n$, and $u_m=\cos\theta_m$ has been substituted for all $m$. 
Using integration by parts for $\int_{-1}^{1}du_m\,u_m\,\exp{(\beta\lambda\,F(u_m))}$ 
and $\int_{-1}^{1}du_m\,\exp{(\beta\lambda\,F(u_m))}\,F'(u_m)$, Eq.~\ref{eq:pq_partition} 
simplifies to 
\begin{equation}
 \label{eq:pq_theo}
 \tilde{A}_n:=\langle{\bf p}_n\cdot{\bf q}_n\rangle = -\frac{6}{\beta \zeta_n}\text{~.}
\end{equation}
Thus, despite the additional angle potential, one has the same short-time value for 
$\langle{\bf p}_n\cdot{\bf q}_n\rangle$ as for a Rouse chain. 
For the terminal monomers, one can show in analogy that the short-time value 
\begin{equation}
 \label{eq:pq_theo_term}
 \tilde{A}_1=\tilde{A}_N:=\langle{\bf p}_n\cdot{\bf q}_n\rangle = -\frac{3}{\beta \zeta_n}
\end{equation}
is also independent of both $k$ and $\lambda$. 

For larger $\Delta n$, one might wonder if the adjacent bonding angles, located at 
monomers $n-1$ and $n+1$, also contribute to the dynamics of monomer $n$. 
However, the net bending forces, resulting from $U_{\theta_{n-1}}$ and $U_{\theta_{n+1}}$, vanish 
since the monomers  $n-2$ and $n+2$ acquire all possible orientations in thermal equilibrium 
(in particular due to the rotation around the bonds ${\bf r}_{n-1}$ and ${\bf r}_{n}$). 
Consequently, for an elementary time step, one also has 
$\langle {\bf p}_n\cdot{\bf q}_n\rangle = -\,6/(\beta\zeta_n)$ for $\Delta n>1$.

\subsection{Numerical Calculation}

\begin{figure}
 \centering
 \includegraphics[scale=0.3]{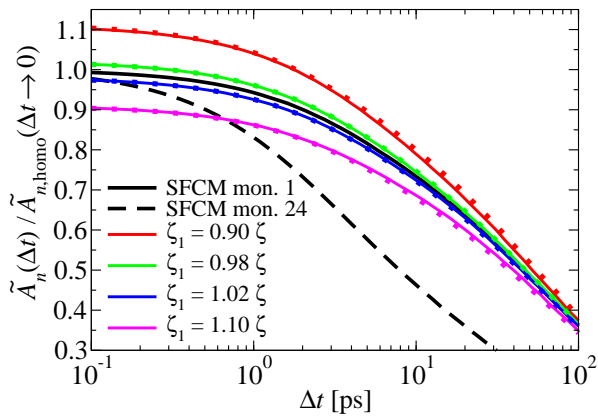}
 \caption{Short-time regime of the correlator $\tilde{A}_n(\Delta t):=\langle {\bf p}_n(\Delta t)\cdot{\bf q}_n\rangle$ 
          for a terminal (black solid line) and a central (black dashed line) SFCM monomer. 
          Additionally, $\zeta_n$ has been changed for specific monomer positions, namely 
          (a)~for the terminal monomers ($n=1$ and $48$, solid colored lines) and (b)~for the four outermost monomers 
          ($n=1-4$ and $45-48$, dotted colored lines). 
          The ordinate has been normalized by the short-time values of the homogeneous SFCM (\ie $\zeta_n=\zeta$). 
          The error bars are smaller than the line thickness. }
 \label{fig:pq_zterm}
\end{figure}

Fig.~\ref{fig:pq_zterm} shows the short-time regime of the correlator 
$\tilde{A}_n(\Delta t):=\langle {\bf p}_n(\Delta t)\cdot{\bf q}_n\rangle$ for a terminal 
(black solid line) and a central (black dashed line) 
monomer of the SFCM as determined from a BD simulation, 
in which a uniform $\zeta$-value has been used. 
In order to convert the model curves to standard units, the time axis has been scaled by the 
ratio of the end-to-end-vector relaxation times $\tau_\mathrm{R}$ of the SFCM and PEO. 
To this purpose, we used a Kohlrausch-Williams-Watts fit for the end-to-end vector 
autocorrelation function, \ie $\langle{\bf R}_\mathrm{e}(0)\cdot{\bf R}_\mathrm{e}(0)\rangle = \langle{\bf R}_\mathrm{e}^2\rangle\exp{(-(t/\tau)^\beta)}$, 
where $\tau_\mathrm{R}$ is given by $\tau_\mathrm{R}=\tau\beta^{-1}\Gamma(\beta^{-1})$, 
and $\Gamma$ is the gamma function. 
The ordinate has been normalized by the short-time values of the 
homogeneous SFCM with $\zeta_n=\zeta$ (\ie Eqs.~\ref{eq:pq_theo} and \ref{eq:pq_theo_term}, respectively). 

As expected, $\tilde{A}_n(\Delta t)$ 
approaches the value $\zeta_n^{-1}$ in the limit $\Delta t\rightarrow 0$. 
On larger time scales, however, the curves decay due to the relaxation of 
the local chain curvature expressed by ${\bf q}_n$. 

In addition, we also calculated $\tilde{A}_1(\Delta t)$ for a SFCM, 
in which (a)~the terminal monomers (\ie $n=1$ and $n=48$) and (b)~the four outermost monomers 
($n=1-4$ and $n=45-48$) had a different friction coefficient $\zeta_1$ (or $\zeta_{\{1-4\}}$, respectively) than the other 
monomers (in particular $\zeta_1=0.9\,\zeta$, $\zeta_1=0.98\,\zeta$, 
$\zeta_1=1.02\,\zeta$ and $\zeta_1=1.1\,\zeta$). 
Also for the model systems with heterogeneous mobilities, the short-time 
value of $\tilde{A}_1(\Delta t)$ can clearly be identified with $\zeta_1^{-1}$ (Fig.~\ref{fig:pq_zterm}). 
Moreover, for a given $\zeta_1$, $\tilde{A}_1(\Delta t)$ is 
nearly the same for the SFCM with $\zeta_1$ at the end monomer only (scenario~a) and the 
SFCM in which the four outermost monomers have a different $\zeta_{\{1-4\}}$-value (scenario~b). 
Thus, also numerically, our method is able to determine $\zeta_n$ on a quantitative level. 

With increasing $\Delta t$, the $\tilde{A}_n(\Delta t)$-curves of all models converge asymptotically towards the same long-time value, 
meaning also that all information about the mobility $\zeta_n^{-1}$ is lost. 
This is an important point, since it turns out that a chemically realistic polymer chain such as PEO does not display a short-time plateau 
because of the microscopic potentials of the backbone and its ballistic dynamics in the sub-picosecond regime 
(these issues are discussed in Section~\ref{sec:pq_application} and also in our previous publications~\cite{DiddensEPL2010,DiddensEPL2011}). 
Due to these effects, the $\zeta_n$ for PEO can only be extracted at larger $\Delta t$. 
In our previous work~\cite{DiddensEPL2010}, we found that the monomer-averaged $\tilde{A}(\Delta t)$-curves for PEO and for the homogeneous SFCM 
agree with each other from about $\Delta t=10\text{~ps}$ on, thus defining a minimum time scale for the determination of the monomeric PEO mobilities. 
At the same time, however, the agreement of $\tilde{A}(\Delta t)$ between SFCM and PEO for $\Delta t\geq 10\text{~ps}$ also allows one to extract the 
PEO mobilities by comparison with the model curves. 
That is, the $\tilde{A}_1$-curves in Fig~\ref{fig:pq_zterm} can be utilized to determine the relative mobility of the terminal and central PEO monomers. 

The remaining issue now is to estimate in how far the $\tilde{A}_1$-curves are still governed by the \emph{local} mobility $\zeta_1^{-1}$ 
for $\Delta t\geq 10\text{~ps}$, since the dynamics of a particular monomer will also be influenced by the other monomers in the segment 
beyond the microscopic time scale. 
Thus, it is a priori uncertain with which accuracy an unknown $\zeta_1$-value can be determined from an observed $\tilde{A}_1$. 
To elucidate this, we plotted $\tilde{A}_1(\Delta t)$ versus the underlying friction coefficient $\zeta_1$ (which is known in case of the model chains) 
for $\Delta t=10\text{~ps}$ and $\Delta t=30\text{~ps}$. 
It turns out that their dependence is linear (see appendix~\ref{sec:appendix_accuracy}). 
In particular, when normalizing both quantities by the respective values for the homogeneous SFCM, 
\ie $\tilde{A}_{1,\mathrm{homo}}$ and $\zeta$, the data can be described by 
\begin{equation}
 \label{eq:translation}
 \frac{\tilde{A}_1(\Delta t)}{\tilde{A}_{1,\mathrm{homo}}(\Delta t)}-1=\rho(\Delta t)\,\left[\frac{\zeta_1}{\zeta}-1\right]\text{~,}
\end{equation}
which provides a translation rule between any observed $\tilde{A}_1$ and the corresponding $\zeta_1$-value. 
For the SFCM in which only the end monomers have a different mobility (scenario~(a), see above), we find $\rho(10\text{~ps})=0.69$ and $\rho(30\text{~ps})=0.55$. 
When the four outermost monomers have different mobilities (scenario~(b)), one has larger values, \ie $\rho(10\text{~ps})=0.82$ and $\rho(30\text{~ps})=0.72$. 
For PEO, however, not only the mobility of the chain ends relative to the central monomers is unknown, 
but also how many successive monomers at the ends might be affected by the different intermolecular environment, 
and therefore might display a different mobility. 
That is, speaking in terms of the SFCM, it is unclear if scenario~(a) or (b) better describes the real physical situation, and thus 
which model is more appropriate to determine $\zeta_1$ from the PEO data. 
Therefore, we tested the robustness of our method by extracting $\zeta_1$ according to Eq.~\ref{eq:translation} of one model using the 
$\rho(\Delta t)$-values of the other model. 
For both combinations, we find that the extracted $\zeta_1$-values deviate by no more than $3\text{~\%}$ from the true values. 
Altogether, this clearly demonstrates that the local mobility of a given monomer can be determined with high precision using the 
SFCM curves from Fig~\ref{fig:pq_zterm}. 

It should be noted, however, that the dynamical behavior beyond the short-time limit also depends on the precise value of the chain 
stiffness (characterized by $\lambda$ for the SFCM, see Eq.~\ref{eq:angle_pot}). 
However, since we already matched the characteristic ratio of the SFCM to that of PEO (section~\ref{sec:sim_details}), the interpretation 
of the PEO data (section~\ref{sec:pq_application}) remains unaffected, and we refer to appendix~\ref{sec:appendix_lambda} for a 
discussion of this effect. 

In principle, when only comparing the model systems, one could obtain the same information from the short-time MSDs (not shown). 
However, for a realistic polymer chain such as PEO, the MSDs in this regime are dominated by dynamical 
contributions arising from the complicated, microscopic potentials (see below), whereas the model-chain 
behavior emerges only on longer time scales, which in turn is mainly governed by the global polymer motion. 
Contrarily, the analysis shown in Fig.~\ref{fig:pq_zterm} is strictly local, as $A_n(\Delta t)$ is only 
sensitive to the relaxation of a given monomer within the local curvature ${\bf q}_n$. 
In other words, whereas all non-ideal dynamical contributions accumulate in the MSD for a realistic 
polymer melt, they act as random noise in the pq-analysis, provided that they are non-systematic with respect to ${\bf q}_n$. 

In summary, the observable $\tilde{A}_n(\Delta t)$ clearly allows the extraction of the \emph{bare} mobility of a given SFCM monomer from its short-time value. 
Contrarily, on time scales larger than about $30\text{~ps}$, $\tilde{A}_n(\Delta t)$ is also governed by the motion of the neighbor monomers, 
and thus rather corresponds to an \emph{effective} mobility. 
In this regime, however, the $\Delta t$-dependence of the SFCM curves also becomes affected by the specific value of $\lambda$ (Eq.~\ref{eq:angle_pot}).

\section{Structural Properties}

\begin{figure}
 \centering
 \includegraphics[scale=0.3]{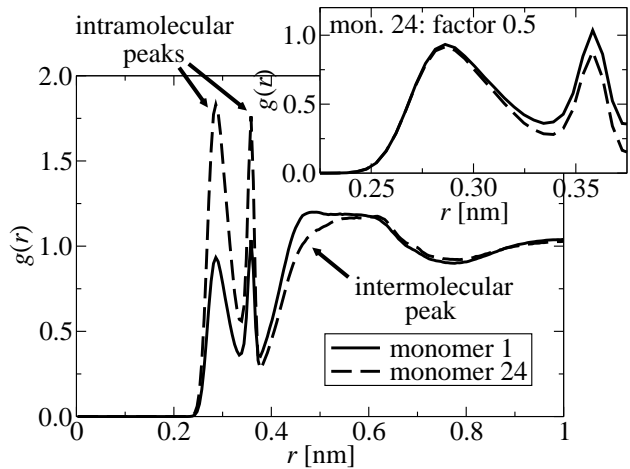}
 \caption{Radial distribution function of the terminal ($n=1$ and $n=48$) and the central ($n=24$ and $n=25$) PEO monomers in the melt. 
          The average was performed over all chains and all initial time frames. 
          For clarity, the curve of the central monomer has been divided by two in the inset. }
 \label{fig:gofr}
\end{figure}

Fig.~\ref{fig:gofr} shows the radial distribution function of the ether oxygens of the PEO chains. 
This quantity has been computed for both the terminal ($n=1$ and $n=48$) and the central ($n=24$ and $n=25$) PEO monomers. 
The average has been performed over all chains in the simulation box as well as over all initial time frames. 
Up to approximately $0.4\text{~nm}$, two intramolecular peaks around $r=0.28\text{~nm}$ and 
$r=0.36\text{~nm}$ can be observed, corresponding to the neighbor monomer(s). 
The occurrence of two peaks indicates the existence of two preferred conformations (note that the 
root mean squared distance between the oxygen atoms of two bonded monomers is about $0.32\text{~nm}$). 
Naturally, these peaks are approximately twice as high for $n=24$ due to chain connectivity (see inset 
of Fig.~\ref{fig:gofr}, where the curve of monomer 24 has been divided by two). 
When going to distances of about $0.4-0.5\text{~nm}$, one indeed observes that the terminal monomer has 
more intermolecular neighbors in its first coordination shell than the central monomer. 

Of course, these observations are not very surprising, since the different intermolecular {\it structure} 
in the vicinity of terminal and central monomers seems only logical. 
However, as pointed out above, exactly these differences may also alter the {\it dynamical} behavior beyond 
the general differences due to the dissimilar chain connectivity, which are already contained in the Rouse model.

\section{Mean Squared Displacement}

\begin{figure}
 \centering
 \includegraphics[scale=0.3]{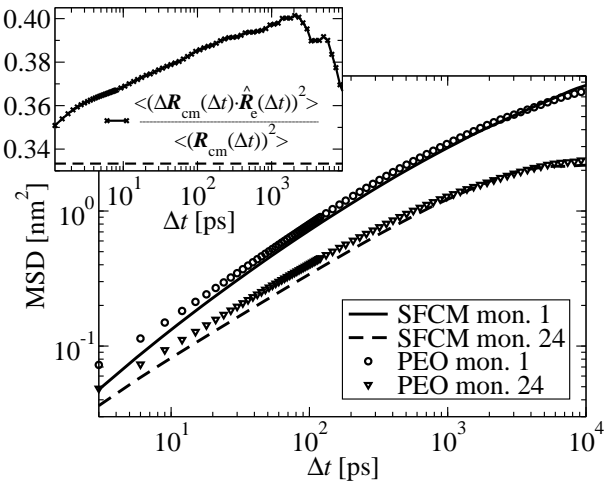}
 \caption{MSDs of the terminal and the central monomers for PEO (symbols) and the SFCM (lines). 
          All curves have been computed in the center-of-mass frame of the chain. 
          The inset shows the ratio of the center-of-mass MSD in direction of the end-to-end vector ${\bf R}_\mathrm{e}$ relative to the total center-of-mass MSD for PEO. 
          The dashed line indicates the ideal value for isotropic diffusion. }
 \label{fig:msd}
\end{figure}

In order to get a first impression of the polymer motion, we start with the MSD, which can be 
regarded as the standard tool to study the dynamics in simulations. 
For the monomeric MSDs, one generally observes that for intermediate time scales 
(\ie $\tau_\mathrm{R}/N^2 \le \Delta t \le \tau_\mathrm{R}$, with $\tau_\mathrm{R}$ being the Rouse time) 
the outer monomers move faster than 
the central segments~\cite{KremerJCP1990,BaschnagelBinderMamol2001,PaulChemPhys2002}. 
Naturally, this regime is already highly affected by the connectivity constraints of the chain, 
which are less present for the end monomers. 
This is also confirmed in Fig.~\ref{fig:msd}, which shows the MSDs of the terminal and the central 
PEO monomers in the center-of-mass frame. 
As stated above, the outer monomers are faster for all $\Delta t$ shorter than $\tau_\mathrm{R}$, 
mainly as a result of the chain connectivity. 
The MSDs of the respective SFCM monomers are plotted as black solid (terminal) and black dashed (central) 
line in Fig.~\ref{fig:msd}. 
For both the central and the outer monomers, one observes that the MSD of the SFCM is lower than 
the PEO curve. 
These deviations can be related to the more complicated local potentials for the latter system. 
When computing the MSD in absolute coordinates, the mismatch between PEO and SFCM becomes even worse 
due to the well-known, non-ideal center-of-mass motion of polymer chains in realistic 
melts~\cite{PaulBinder1991,PaulPRL1998,PaulSmithRichterChemPhys2000,PaulSmithRev2004,WittmerBaschnagelJCP2011}, which 
is subtracted in Fig.~\ref{fig:msd}. 
For these reasons, it is difficult to judge solely from the monomeric MSDs whether the Rouse assumption 
of a uniform $\zeta$ also holds for the PEO melt. 

\section{Application of the pq-Method}
\label{sec:pq_application}

\begin{figure}
 \centering
 \includegraphics[scale=0.3]{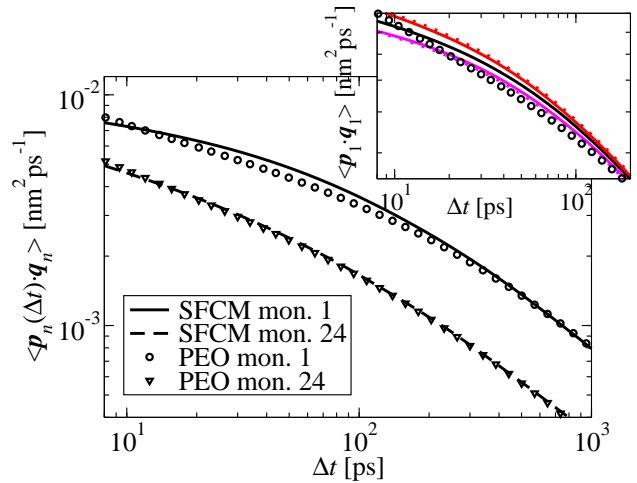}
 \caption{Correlator $\tilde{A}_n(\Delta t)$ as a function of $\Delta t$ for PEO (symbols) and for the SFCM (lines) for terminal and central monomers.  
          The inset shows the curve for the terminal PEO monomer together with the model curves from Fig.~\ref{fig:pq_zterm} in the range $\Delta t=8-200\text{~ps}$. 
          For all curves, the error bars are smaller than the line thickness/symbol size. }
 \label{fig:pq_SFCM_vs_PEO}
\end{figure}

Alternatively, the local dynamics of the individual PEO monomers can be evaluated by the pq-method, yielding an effective mobility free from connectivity effects. 
The $\tilde{A}_n(\Delta t)$-curves for the terminal (circles) and the central (triangles) PEO monomers as well as the corresponding curves for the SFCM with a uniform 
$\zeta$-value (black lines) are shown in Fig.~\ref{fig:pq_SFCM_vs_PEO}. 
As above, the $\Delta t$-axis of the model curves has been scaled by the ratio of the $\tau_\mathrm{R}$-values of PEO and the SFCM. 
The $\tilde{A}_n(\Delta t)$-axis has been scaled by the ratio of the mean squared Kuhn lengths $b^2$ and the $\tau_\mathrm{R}$ ratio. 

In case of PEO, we notice that, unlike for the SFCM, no short-time plateau emerges on time scales shorter than $10\text{~ps}$ (not shown in Fig.~\ref{fig:pq_SFCM_vs_PEO}). 
Rather, $\tilde{A}_n(\Delta t)$ increases steadily with decreasing $\Delta t$, thus yielding no direct access to the monomers' mobilities. 
Of course, this behavior is not too surprising, since it seems likely that the chemical potentials and the excluded volume interactions give rise to 
additional systematic contributions to the short-time dynamics of PEO. 
However, for $\Delta t\ge 10\text{~ps}$, one observes reasonable agreement between PEO and the SFCM for both monomer positions, demonstrating that the 
SFCM basically captures the local dynamical features of the PEO chains in the melt on all except shortest time scales. 
In particular for $\Delta t\approx 10\text{~ps}$, for which the SFCM dynamics is still mainly governed by $\zeta_n$ 
(see discussion in context with Fig.~\ref{fig:pq_zterm}), the agreement is quantitative. 
This demonstrates that the short-time mobilities of the individual PEO monomers are uniformly distributed, thus confirming the classical Rouse assumption. 

Apart from the short-time deviations between PEO and the SFCM, minor deviations also become 
noticeable for the terminal monomers on intermediate time scales, \ie $\Delta t\approx 20-200\text{~ps}$. 
This can also be seen in the inset of Fig.~\ref{fig:pq_SFCM_vs_PEO}, which shows the curve of 
the terminal PEO monomer in the time window of $8-200\text{~ps}$ together with the model curves 
from Fig.~\ref{fig:pq_zterm} (\ie $\zeta_1=0.9\,\zeta$ and $\zeta_1=1.1\,\zeta$). 
It seems that the dynamics of the terminal PEO monomer on this time scale rather corresponds to that of an SFCM monomer with 
$\zeta_1\approx 1.1\,\zeta$, although it is unclear whether this apparent agreement is just by mere coincidence. 
Especially due to the fact that $\tilde{A}_1(\Delta t)$ for PEO agrees with the SFCM at $\Delta t\approx 10\text{~ps}$, 
and also the central and the monomer-averaged $\tilde{A}(\Delta t)$-curves agree for \emph{all} $\Delta t\geq 10\text{~ps}$, 
the naive interpretation $\zeta_1\approx 1.1\,\zeta$ for $\Delta t\approx 20-200\text{~ps}$ has to be considered carefully. 
Within this context, it is also important to bear in mind that the dynamics on the time scale of 
$\Delta t\approx 20-200\text{~ps}$ is not solely determined by $\zeta_n^{-1}$, but also by 
the motion of more remote monomers, and, more importantly, the details of the chemical potentials 
(cf. discussion of Fig.~\ref{fig:pq_zterm}). 
Of course, in contrast to the effective-medium picture of the Rouse model or the SFCM derived thereof, 
the motion of the PEO monomers is not only affected by the chain connectivity, but also by intra- and 
intermolecular excluded volume interactions. 

A first indication that these two effects (\ie chain connectivity and correlated motion of neighbors) 
have a different impact on a terminal and on a central monomer is given by the following observation: 
The average time during which a terminal monomer diffuses a distance approximately equal to the size 
of its own coordination sphere (\ie $r_\mathrm{c}=0.78\text{~nm}$, as estimated from the extent of the 
first intermolecular peak in Fig.~\ref{fig:gofr}) is about $\Delta t\approx66\text{~ps}$ (Fig.~\ref{fig:msd}), 
thus falling exactly into the range of $20-200\text{~ps}$ where the deviations in Fig.~\ref{fig:pq_SFCM_vs_PEO} occur. 
Contrarily, the average central monomer has not diffused this far until $200\text{~ps}$. 
That is, the coordination sphere of a terminal monomer itself, mostly consisting of central monomers which 
are intrinsically slower due to chain connectivity, relaxes on larger time scales than the self-motion of 
the outer monomers. 
In case of the central monomers, the intermolecular surroundings (modeled as frictious background in the 
Rouse model) relax on the same time scale as the tagged monomer. 

\begin{figure}
 \centering
 \includegraphics[scale=0.2]{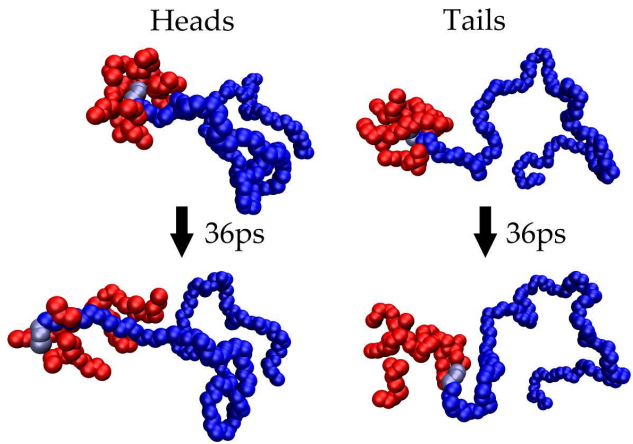}
 \caption{Snapshots from the MD simulations for two events which have been identified as head (left) or 
          tail monomer (right) at $\Delta t=36\text{~ps}$. 
          The tagged chain is shown in blue, whereas the surrounding monomers in the first coordination 
          sphere of the initial configuration are shown in red. 
          The terminal monomer is shown in light blue. 
          The snapshots have been created with VMD~\cite{VMD}. }
 \label{fig:snapshots}
\end{figure}

Another fundamental difference is that a terminal monomer may exit its initial coordination sphere either 
by moving away from the center of the polymer chain, thus paving the way for the remaining part of the 
PEO molecule (subsequently termed as head monomers), or by following the contour of the rest of the chain 
(tail monomers, see snapshots from the MD simulations in Fig.~\ref{fig:snapshots}, where these two events 
have been identified). 
In contrast to this, there is no such asymmetry for a monomer located in the center of the chain, as these 
monomers experience backdragging forces on both sides. 

\begin{figure}
 \centering
 \includegraphics[scale=0.3]{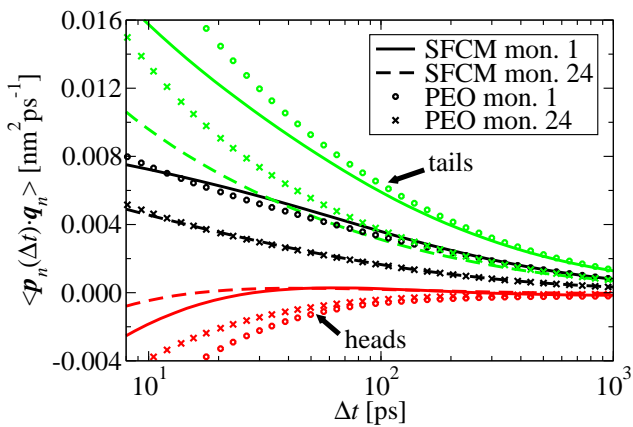}
 \caption{Correlator $\tilde{A}_n(\Delta t)$ for the terminal and the central monomers of PEO and the SFCM. 
          Here, a distinction was made in which direction (relative to ${\bf q}_1$ or ${\bf q}_{24}$) the center of mass of the subchain 
          consisting of all monomers within ${\bf q}_n$ moved during $\Delta t$. }
 \label{fig:pq_head_tail}
\end{figure}

In order to investigate these directional correlations in more detail, we modified our analysis in the following way: 
For a given displacement 
$\Delta {\bf R}_\mathrm{cm}^{(1-5)}(\Delta t)={\bf R}_\mathrm{cm}^{(1-5)}(t+\Delta t)-{\bf R}_\mathrm{cm}^{(1-5)}(t)$ 
of the terminal Kuhn segment (monomers $1-5$, with center-of-mass position ${\bf R}_\mathrm{cm}^{(1-5)}$), 
a distinction was made whether it moved in or against the direction of the ${\bf q}_1$-vector. 
In this way, one can distinguish between head monomers 
($\Delta {\bf R}_\mathrm{cm}^{(1-5)}(\Delta t)\cdot\hat{{\bf q}}_1(t)>0$), where the terminal Kuhn 
segment moves in front of the adjacent segments, and tail monomers 
($\Delta {\bf R}_\mathrm{cm}^{(1-5)}(\Delta t)\cdot\hat{{\bf q}}_1(t)<0$), where this segment follows 
the local chain contour. 
These two contributions to $\tilde{A}_1$ and to $\tilde{A}_{24}$ are shown in Fig.~\ref{fig:pq_head_tail} 
(see also snapshots from the MD simulations in Fig.~\ref{fig:snapshots}). 
For $n=24$, the criterion to define head and tail monomers was if the center of mass of the subchain defined by the monomers $n-\Delta n, \, \ldots, \, n$, or, alternatively, 
$n, \, \ldots, \, n+\Delta n$ moved in or against the direction of the bond vector of the respective Kuhn segment (see also sketch in Fig.~\ref{fig:sketch}). 
Note that in both cases the curves for the head monomers have also negative values, which result from their head-monomer definition. 
Naturally, trivial head and tail contributions can also be observed for the SFCM (Fig.~\ref{fig:pq_head_tail}) due to the bias resulting from the distinction between heads and tails. 
Interestingly, despite the good agreement of the average curves (in particular for $n=24$), the absolute values of both head and tail contributions for PEO are larger than for the SFCM. 
This is a consequence of the additional excluded volume, since for PEO all monomers within one Kuhn segment move much more correlated. 
Thus, the short-time displacement of a PEO monomer will already be significantly influenced by the motion of the other monomers in the segment, whereas the SFCM monomers can interpenetrate each 
other and thus exhibit weaker motional correlations (which even vanish in the limit $\Delta t\rightarrow 0$). 
Remarkably, for both the terminal and the central monomers, the head and tail contributions roughly cancel each other, 
and the local PEO dynamics in the melt is essentially the same as for a phantom chain (with minor deviations for the terminal monomers). 

Of course, for PEO, one might wonder if the different coordination sphere of the terminal monomers 
affects the surrounding monomers themselves. 
For instance, it has recently been found for melts consisting of poly(propylene oxide) oligomers that 
the average polymer relaxation is faster compared to melts of longer chains (\ie $N\ge 20$), which 
could be attributed to the larger free volume of the chain ends\cite{VogelMamol2010}. 
In analogy, one might expect a similar effect for the local mobility. 
To this purpose, we calculated $\tilde{A}_n(\Delta t)$ for all intermolecular 
neighbors of the terminal monomers (\ie up to a maximum distance of $0.5\text{~nm}$, cf. Fig~\ref{fig:gofr}). 
In total, we only observe a marginal increase of $\tilde{A}_n(\Delta t)$ 
of up to $5\text{~\%}$, which is most pronounced for time scales below $10\text{~ps}$ and becomes 
negligible for larger $\Delta t$ (not shown). 
Thus, also the monomers in the vicinity of a chain end essentially display Rouse-like motion 
on local scales. 

Finally, the enhanced correlations of the monomer dynamics along the PEO backbone also manifest themselves in the MSD of the center of mass of the PEO chains. 
Apart from the well-known subdiffusivity on time scales shorter than $\tau_\mathrm{R}$, which has 
been found in several simulations~\cite{PaulBinder1991,PaulPRL1998,PaulSmithRichterChemPhys2000,PaulSmithRev2004,WittmerBaschnagelJCP2011}, 
experiments~\cite{PaulSmithRichterChemPhys2000,GuenzaRichterJPCB2008} and theoretical 
analyses~\cite{SchweizerJCP1989,GuenzaPRL2001,WittmerBaschnagelJCP2011,FaragoPRL2011,FaragoJPhysCondensMatter2012}, 
we additionally observe that the center-of-mass diffusion is slightly anisotropic with respect to 
the orientation of the end-to-end vector ${\bf R}_\mathrm{e}$. 
This is demonstrated by the inset of Fig.~\ref{fig:msd}, which shows the ratio of the component 
parallel to ${\bf R}_\mathrm{e}$ relative to the total center-of-mass MSD, \ie 
$\langle[({\bf R}_\mathrm{cm}(t+\Delta t)-{\bf R}_\mathrm{cm}(t))\cdot\hat{{\bf R}}_\mathrm{e}(t)]^2\rangle/\langle[{\bf R}_\mathrm{cm}(t+\Delta t)-{\bf R}_\mathrm{cm}(t)]^2\rangle$. 
In the absence of anisotropies one would expect an ideal ratio of $1/3$. 
However, the actual ratio is larger for all $\Delta t$ shorter than the Rouse time 
($\tau_\mathrm{R}\approx 8.5\text{~ns}$ for PEO). 
This shows that the preferential motion is along the primary axis of the chain, \ie parallel to 
${\bf R}_\mathrm{e}$. 
In how far the observed anisotropy is related to the fact that real polymer coils are not spherical, 
but rather stretched in direction of 
${\bf R}_\mathrm{e}$~\cite{MaassJCP2001,BaschnagelAdvCollInterfSci2001} might be investigated more 
thoroughly in future work. 

In addition to these correlations, memory effects may come into play, which could cause additional 
differences between head and tail monomers. 
That is, when a monomer moves either direction, the surrounding monomers would occupy the newly 
available space in this scenario, and thus block the backward motion.

\section{Conclusion}

In this contribution, we checked the fundamental Rouse assumption of a uniform friction coefficient on all monomers for a PEO melt. 
The mobilities were extracted from MD simulations using our previously developed pq-method~\cite{DiddensEPL2010}, which avoids the classical mode picture and rather employs a 
Langevin-like equation to characterize the local polymer dynamics. 
In contrast to the MSD, this procedure leads to the cancellation of the non-trivial terms for PEO, which arise from the additional chemical potentials. 
In order to interpret the local PEO dynamics, we used a semiflexible phantom chain as a reference. 

During the course of our analysis it turned out that the effective mobility of both the terminal and the central PEO monomers is essentially the same as for the SFCM for 
$\Delta t\approx 10\text{~ps}$. 
However, this agreement results from the nearly quantitative cancellation of the more complicated interactions in the PEO melt. 
A more detailed analysis revealed that the relaxation with respect to the local chain curvature expressed by ${\bf q}_n$ can be decomposed in two individual contributions 
(head and tail monomers), depending on the direction of motion of the Kuhn segment under consideration. 
Due to the correlated motion in PEO arising from the excluded volume interactions, both head and tail contributions are larger for PEO. 
Remarkably, these contributions approximately cancel, and the mobility is roughly the same for all PEO monomers. 
Minor deviations (up to ten percent) only become noticeable on intermediate time scales ($\Delta t\approx 20-200\text{~ps}$) for the terminal monomers, for which the head 
and tail dynamics is not entirely equal. 

As reported previously~\cite{DiddensEPL2011}, our findings clearly demonstrate that the pq-method yields complementary information on the dynamics of macromolecular systems. 
In further work one might study in how far the chain end effects persist or enhance when approaching the glass transition temperature (cf. the Flory-Fox equation). 
Moreover, the pq-method is also supposed to yield fruitful results for confined polymer melts~\cite{BaschnagelPRE2002,BaschnagelJPolymSci2006} or complex polymer architectures~\cite{SommerMacromol2007}.

\begin{acknowledgments}
The authors would like to thank J\"org Baschnagel, Hendrik Meyer, Jean Farago and Micheal Vogel 
for helpful discussions and correspondence. 
Financial support from the NRW Graduate School of Chemistry is also greatly appreciated. 
\end{acknowledgments}

\appendix

\section{Relation between $\tilde{A}_n(\Delta t)$ and mobility $\zeta_n^{-1}$}
\label{sec:appendix_accuracy}

\begin{figure}[h]
 \centering
 \includegraphics[scale=0.32]{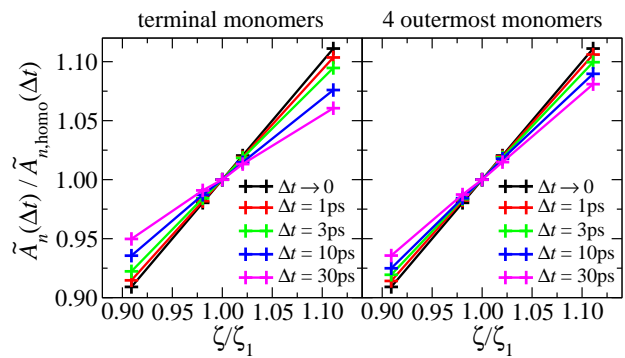}
 \caption{Normalized $\tilde{A}_1$-values in dependence of the underlying $\zeta_1$ employed in the BD simulations for various $\Delta t$. }
 \label{fig:zterm_accuracy}
\end{figure}

In order to estimate in how far the $\tilde{A}_1$-curves are governed by the local mobility $\zeta_1^{-1}$ 
for larger $\Delta t$-values, Fig.~\ref{fig:zterm_accuracy} shows $\tilde{A}_1(\Delta t)$ as a function of the underlying 
friction coefficient $\zeta_1$ (which is known in case for the SFCM) at $\Delta t\rightarrow 0$, $\Delta t=1\text{~ps}$, 
$\Delta t=3\text{~ps}$, $\Delta t=10\text{~ps}$ and $\Delta t=30\text{~ps}$. 
By normalizing both quantities by the respective values for the homogeneous SFCM, \ie $\tilde{A}_{1,\mathrm{homo}}$ and $\zeta_1=\zeta$, 
the data can be described by Eq.~\ref{eq:translation}. 
The resulting $\rho(\Delta t)$-values are presented in Table~\ref{tab:zterm_accuracy}. 

\begin{table}[h]
 \centering
 \begin{tabular}{r c c c}
  \hline
  \hline
   & \multicolumn{3}{c}{\hspace{-0.8cm} $\rho(\Delta t)$} \\
  $\Delta t$ & terminal monomers & \hspace{0.2cm} & four outermost monomers \\
  \hline
  $\Delta t\rightarrow 0$ & $1.00$ &  & $1.00$ \\
  $1\text{~ps}$ & $0.93$ &  & $0.95$ \\
  $3\text{~ps}$ & $0.85$ &  & $0.89$ \\
  $10\text{~ps}$ & $0.69$ &  & $0.82$ \\
  $30\text{~ps}$ & $0.55$ &  & $0.72$ \\
  \hline
  \hline
 \end{tabular}
 \caption{Fitting parameter $\rho$ (Eq.~\ref{eq:translation}) for the SFCM models as a 
          function of $\Delta t$ as extracted from Fig.~\ref{fig:zterm_accuracy}. }
 \label{tab:zterm_accuracy}
\end{table}

\section{Influence of the chain stiffness $\lambda$ on $\tilde{A}_n(\Delta t)$}
\label{sec:appendix_lambda}

\begin{figure}[h]
 \centering
 \includegraphics[scale=0.3]{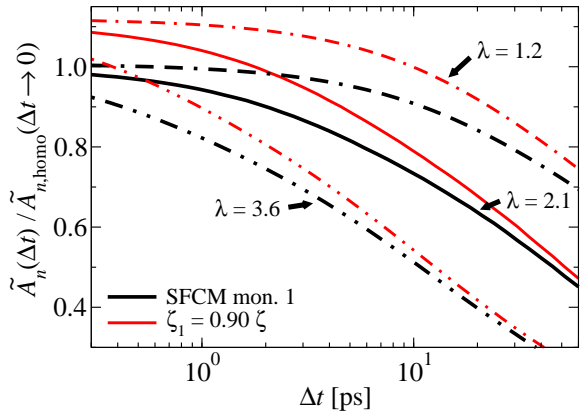}
 \caption{Correlator $\tilde{A}_n(\Delta t)$ for the homogeneous SFCM and the SFCM with $\zeta_1=0.9\,\zeta$ 
          for different lambda values (in units of $k_\mathrm{B}T$, Eq.~\ref{eq:angle_pot}). }
 \label{fig:pq_lambda}
\end{figure}

Fig.~\ref{fig:pq_lambda} shows $\tilde{A}_1(\Delta t)$ for the homogeneous SFCM and for an SFCM in which the terminal monomers have a friction coefficient of $\zeta_1=0.9\,\zeta$ 
for two additional values of $\lambda$ (\ie $\beta\lambda=1.2$ and $\beta\lambda=3.6$, Eq.~\ref{eq:angle_pot}). 
One clearly observes that the different stiffness gives rise to a quantitatively different decay of $\tilde{A}_1(\Delta t)$, and only the asymptotic short-time limit remains unaffected. 
Thus, when comparing the dynamics of the SFCM with another polymer species such as PEO on larger time scales (section~\ref{sec:pq_application}), one has to adjust the stiffness of 
the model chain, as \eg done in the present work by reproducing the $C_\infty$-value. 

\bibliography{literatur}

\end{document}